\documentstyle[epsf,twoside,a4wide,12pt]{article}
\def\IJMP#1{Int. J. Mod. Phys.~{\bf #1}}

\def\NP#1{Nucl. Phys.~{\bf #1}}

\def\PL#1{Phys. Lett.~{\bf #1}}
\def\PR#1{Phys. Rev.~{\bf #1}}
\def\PRP#1{Phys. Rep.~{\bf #1}}

\def\RMP#1{Rev. Mod. Phys.~{\bf #1}}

\newcommand\D{\displaystyle}

\newcommand\be{\begin{equation}}
\newcommand\ee{\end{equation}}
\newcommand\bano{\begin{eqnarray*}}
\newcommand\eano{\end{eqnarray*}}
\newcommand\ba{\begin{eqnarray}}
\newcommand\ea{\end{eqnarray}}

\begin{document}

\title{\vskip -1cm\hfill {\large HD-TVP-97-15}\\
\vskip 1cm
Nonperturbative Flow Equations\\with Heat-Kernel Methods\\
at finite Temperature
\thanks{Talk given by the first author 
at Research Workshop on Deconfinement at Finite Temperature 
and Density, JINR Dubna, Russia, October 1-29, 1997}
\thanks{supported by GSI, Darmstadt}} 
\author{Bernd-Jochen Schaefer
\thanks{e-mail address: schaefer@hybrid.tphys.uni-heidelberg.de}
\\and\\
Hans-J\"urgen Pirner
\thanks{e-mail address: pir@dxnhd1.mpi-hd.mpg.de}\\
\\
Institut f\"ur Theoretische Physik\\Universit\"at Heidelberg\\
}

\vspace{1.0cm}

\maketitle

\begin{abstract}
\setlength{\baselineskip}{18pt}
\noindent
We derive nonperturbative flow equations within an ef\-fect\-ive
con\-stitu\-ent quark model for two quark flavors. 
Heat-kernel methods are employed 
for a renormalization group improved effective potential.
We study the evolution of the effective potential 
with respect to an infrared cutoff scale
$k$ at vanishing temperature. At the first stage we omit 
corrections coming from the anomalous dimension.
This investigation is extrapolated
to finite temperature, where we find a second order phase transition 
in the chiral limit at $T_c \approx 130$ MeV. 
Due to a smooth decoupling of massive modes, we can directly link
the low-temperature four-dimensional theory to the three-dimensional
high-temperature theory and can determine universal critical 
exponents.
\end{abstract} 
\newpage \parskip12pt

\section{Introduction}

The linear $\sigma$-model has become one of the most used working tools for 
the investigation of the chiral phase transition in QCD\@. It shares an
adequate realism with good practicality in applications. Besides the study
of equilibrium phenomena, the  $\sigma$-model has also been widely applied 
to calculations of the disoriented chiral condensate. In a series of 
ground-breaking papers Wetterich and his group \cite{wett,dirk} have 
applied renormalization group methods to calculate the parameters of the
$\sigma$-model for all resolution scales.
The fundamental idea is to follow the dynamics of the system by integrating
out quantum fluctuations in infinitesimal 
intervals from a high momentum scale to the
far infrared \cite{block,polc}.
In practice these exact renormalization group flow equations 
have to be truncated in a similar way to Schwinger-Dyson equations.
They allow to treat 
the critical fluctuations of the long range $\sigma$- and
$\vec{\pi}$-fields near the second order phase transition, where mean field 
methods or sub summations of the effective mass type are insufficient.
This approach also has to make approximations on the form
of the effective action. In the following paper we follow the
spirit of this renormalization group approach, parameterizing the
shape of the effective potential. In fact we only allow  quartic and 
quadratic couplings in the effective mesonic potential. 
This approximation together with a novel heat-kernel infrared cutoff
prescription allows us to derive new and very transparent formulae for the
evolution equations. 
Since our cutoff function is different from previous work
it allows to see effects of 
such a variant of the cutoff functions used in Wetterichs group 
\cite{wett,dirk}.
The main aim of our contribution is to give  analytical insight
into the physics inherent in the method of evolution equations.

For high resolution, i.e.\ short distance processes fundamental QCD
with quarks and gluons  is the most efficient
theory. A typical scale
associated with perturbative QCD is $\Lambda \geq 1.5$ GeV.
RHIC and LHC physics for secondary particles with $p_{\perp} \geq \Lambda$
will be dominated by such processes. In nucleus nucleus collisions, however,
scattering at smaller momentum scale will be non negligible. Since the
strength of the QCD interactions increases with decreasing momentum transfer,
characteristic $\bar q q$ bound states will form and influence the dynamics
at larger distances. A typical resolution where these processes start
to become important
is $\Lambda_{\chi SB} \approx 1.0$ GeV.
Below this scale the vacuum changes and acquires a quark condensate and/or
meson condensate. 
In a recent paper on deep inelastic scattering \cite{pirn}
it has been shown that 
a photon with varying resolution $Q^2$ is a good physical probe 
to see the transition from partons to constituent
quarks experimentally. Around $Q^2=1$ GeV$^2$ the behavior of the 
structure function $F_2$ changes qualitatively 
as a function of $Q^2$ indicating 
that nature knows about the phenomenon of chiral symmetry
restoration. We think that our heat-kernel
cutoff  restricting the virtualities of the intermediate states
corresponds to the resolution scale of the photon. 
We use the experimental indication for the transition at $1$ GeV
as input to our calculation.
In the  spontaneously
broken phase of chiral symmetry the constituent quark mass will increase 
with decreasing resolution and 
remain finite
down to $\Lambda \approx
\Lambda_{QCD}$. Around this scale the confining gluon 
configurations make themselves
felt via confining forces.
In the interval $\Lambda_{QCD} \leq k \leq \Lambda_{\chi SB}$ the dynamics is
governed by constituent quarks interacting via pions and $\sigma$-mesons.
One sees nicely in our approach
how the different quantum fluctuations from
the $\sigma$-mesons and constituent 
quarks become unimportant when the infrared scale 
parameter becomes smaller than the respective masses
of these states. These modes then decouple
from the further evolution, leaving the zero mass pions alone
in the evolution.

After  having solved the evolution equations at zero temperature for
reasonable starting values of the coupling constants,
we  pursue the evolution at finite 
temperature.
Here the relevant parameter is the
ratio of the temperature over the infrared scale
parameter. 
Decoupling now sets in when the ratio of masses plus the
Matsubara frequencies over the
infrared scale becomes large.
At high temperatures the summation over Matsubara frequencies
is dominated by the lowest mode, thereby reducing the
dynamics to the corresponding three-dimensional field
theory, which is the purely bosonic $O(4)$-model.
We find for the critical index $\beta=0.40$ which is in good
agreement with the results from Monte Carlo calculations \cite{mc}, 
$\epsilon$-expansions
\cite{eps} and of \cite{dirk,morri}.

The outline of the paper is as follows: In section 2 we derive the
evolution equations assuming a fixed parameterization of
the effective potential. In  section 3 we give  the explicit 
expressions of the differential equations for the broken and symmetric
phase both for zero and finite temperature. Section 4 is devoted to
a presentation and  discussion of the results 
obtained after numerical integration of the
evolution equations.

\section{Effective action with infrared cutoff}

In this section we show the derivation of the flow equation 
for the $SU(2) \times SU(2)$ constituent quark model (CQM).

The partition function for the $SU(2) \times SU(2)$ model at zero 
temperature is  given by
\ba
Z[ J=0 ] & = & \int {\cal D}q{\cal D}\bar{q} {\cal D}\sigma {\cal D}\vec{\pi}
\exp \{- \int d^4 x ( {\cal L}_F + {\cal L}_B ) \}\quad.
\ea
We omit external sources $J$ and investigate the chiral limit.
The Euclidean space lagrangian in $d=4$ dimensions looks like
\ba
{\cal L}_F & = & \bar{q}(x) \left( \gamma_E \partial_E +
g\left( \sigma + i\vec{\tau}\vec{\pi}\gamma_5 \right)\right) q(x)\quad,\\
{\cal L}_B & = & \frac{1}{2} \left(
(\partial_\mu \sigma)^2 + (\partial_\mu \vec{\pi})^2 \right) +
\frac{m_0^2}{2}(\sigma^2+\vec{\pi}^2)
+\frac{\lambda_0}{4}(\sigma^2+\vec{\pi}^2)^2\quad.
\ea

The $T=0$ parameters of the linear $\sigma$-model
are fixed at the ultraviolet scale $\Lambda=1.2$ GeV 
in a similar way  as in ref. \cite{dirk}.
At this scale the
quarks are massless partons and the
$\sigma$-field has no vacuum expectation value.
The mass squared  $m_0^2=(0.550$ GeV$)^2$
is positive reflecting
a symmetric ground state.
The minimum of the effective potential
$U(\sigma)$
lies at  the origin.
The other couplings are chosen as
$\lambda_0 = 40.0$ and  $g= M_q / f_{\pi} = 3.23$,
where $g$ is the Yukawa coupling of the constituent quarks to the
mesons and its value corresponds to a constituent quark mass $M_q = 300$ MeV.
We do not evolve the quark Yukawa coupling in this 
introductory work. The result of the evolution will
be largely insensitive to the exact starting value of $\lambda_0$.
We could also have chosen $\lambda_0=0$ 
corresponding  to an initial Lagrangian of the Nambu--Jona-Lasinio
type without explicit mesonic interactions and non-propagating 
mesons \cite{eguc,eber}.

Integration over the fermions yields formally a non-local determinant, which
can be defined by a heat-kernel representation: 
\ba
Z[ J=0 ] & = & \int  {\cal D}\sigma {\cal D}\vec{\pi}
\det \left(\gamma_E \partial_E + g M(x) \right)
\exp \{- \int d^4 x  {\cal L}_B  \}\nonumber\\
& = & \int  {\cal D}\sigma {\cal D}\vec{\pi}
\exp \{ -\frac{1}{2} Tr \log DD^+ -\int d^4 x  {\cal L}_B  \}\\
& = & \int  {\cal D}\sigma {\cal D}\vec{\pi}
\exp \{- \int d^4 x {\cal L}_B + \nonumber\\
&& \frac{1}{2} \int_{1/\Lambda^2}^{\infty} \frac{d\tau}{\tau}
\int d^4 x\; tr \langle x| e^{-\tau \left( -\partial_E^2 +g^2 MM^+ +
g \gamma\cdot(\partial M^+) \right)} | x \rangle \}\quad,\nonumber
\ea 
where we use the abbreviations 
$M(x) = \sigma (x) + i\vec{\tau}\vec{\pi} (x) \gamma_5$ and
$D = \gamma_E \partial_E + g M(x)$.
We remark that the lower boundary of the
integral over the proper time $\tau$ reflects the
ultraviolet scale $\Lambda$ which is fixed
during the whole calculation. In fact, we will choose our 
infrared cutoff function in such a way that no
ultraviolet divergences arise.
Details concerning the heat-kernel representation 
and definitions can be found in ref.~\cite{bjhj}.

In the following we are not interested in wave function renormalizations.
This means that we set the wave function renormalization constant $Z_k = 1$.
We rewrite the meson fields in vectorial form with
$\vec{\phi}=(\sigma,\vec{\pi})$ and 
$MM^+=\vec{\phi}^2$.
We omit all derivatives in the heat-kernel expression and get
for the partition function
\ba
Z[ J=0 ]  = \nonumber 
\ea
\ba
\int  {\cal D}\sigma {\cal D}\vec{\pi}
\exp \{ \frac{1}{2} \int_{1/\Lambda^2}^{\infty}
\frac{d\tau}{\tau}
\int d^4 x\, tr \langle x| e^{-\tau \left( -\partial_E^2 +g^2 \vec{\phi}^2 
\right)} | x \rangle -\int d^4 x  {\cal L}_B  \}\qquad. 
\ea

In principle, the fermion determinant is part of the functional
integration over the meson fields weighted by the meson lagrangian
${\cal L}_B$. In the following, however, we will limit the
integration over fermion and boson fluctuations to the one-loop level
with the additional condition  that the virtualities of the
propagators in the loops are restricted to the interval $\left[k^2,\Lambda^2
\right]$,
where $k^2$ is the infrared and $\Lambda^2$ the ultraviolet cutoff,
i.e.\ we do not consider the effect of modified 
fermions on the meson dynamics in agreement with the  one-loop 
approximation.
The total effective action governing the dynamics of the slowly varying
meson fields $\tilde{\vec{\phi}}$ with virtualities $\leq k^2$ becomes a sum 
of fermionic and bosonic terms.

Using a plane wave basis for the diagonal part of the 
heat-kernel the effective action for the fermions $\Gamma^F$
can be defined by
\ba
\Gamma^F (\tilde{\phi} )& = & \frac{1}{2}\int d^4 x
\int_{1/\Lambda^2}^\infty 
\frac{d\tau}{\tau}
\int \frac{d^4 q}{(2\pi)^4} \left\{ tr_{{N_c}{N_f}\gamma}
e^{-\tau (q^2 + g^2 \tilde{\vec{\phi}^2})} \right\}\quad.
\ea
Here the remaining trace goes over color-, flavor- and spin-space.
We leave out gradients of $\tilde{\phi}$ in the exponential and restrict the
background field $\tilde{\phi}$ to low virtualities. The equivalent one-loop 
integration over the meson fields can be done by expanding the mesonic 
potential $V_0=\frac{m_0^2}{2}(\vec{\phi}^2)
+\frac{\lambda_0}{4}(\vec{\phi}^2)^2$ around the 
slowly varying field configurations 
and one finds for the effective action for the bosons
\ba
\Gamma^B (\tilde{\phi}) & = & -\frac{1}{2}\int d^4 x
\int_{1/\Lambda^2}^\infty 
\frac{d\tau}{\tau}
\int \frac{d^4 q}{(2\pi)^4} \left\{ tr_N 
e^{-\tau (q^2 + \frac{\partial^2 V_0}{\partial \phi_i \partial
    \phi_j})} \right\}\quad.
\ea
Here the trace goes over the $(4 \times 4)$ fluctuation matrices in
$(\sigma,\vec{\pi})$-space. Note the opposite sign of the bosonic and fermionic
actions. The total effective action in one-loop approximation is given by the
sum of both actions:
\ba
\Gamma(\tilde{\phi})=\Gamma^F (\tilde{\phi})+\Gamma^B (\tilde{\phi}).
\ea
In order to implement in this effective action the infrared cutoff on
virtualities one introduces a universal
$k$-dependent
function $f_k ( x = \tau k^2 )$ into the proper time integrand. 
Note, that  the proper time gives the inverse of the virtualities
of the system. We mentioned this already in the discussion of
the ultraviolet cutoff. The function $f_k$ has to
satisfy some general conditions: Since the
action $\Gamma_k$ with infrared cutoff  should tend to the
effective action $\Gamma$ at $k=0$, one must require that
\ba
f_k ( 0 ) & = & 1\quad.
\ea
To suppress modes with small virtualities
the function $f_k$  must also satisfy
\ba
f_k ( x \to \infty ) \to 0\quad.
\ea
This condition regularizes the infrared region of the effective action.
In addition one needs a further condition on the first derivative of the
cutoff function $f_k( x )$
in order to enforce that the flow equation for $\Gamma_k$ is ultraviolet
finite:
\ba
f_k ' (x) & = & -x^2 g(x)
\ea
with $g(x)$ being a regular function in the vicinity of the origin.

One possible choice for the function $f_k (x)$ which fulfills all
these requirements is
\ba
f_k (x) & = & e^{-x} ( 1 + x + \frac{1}{2} x^2 )\quad.
\ea     

The effective action with virtuality cutoff function $f_k (\tau k^2)$
has the following form:
\ba
\Gamma_k [\tilde{\phi}] & = & \int d^4 x V_k (\tilde{\phi})
\ea
with
\ba\label{potential}
V_k (\tilde{\phi})  =  - \frac{1}{2} \int_{0}^\infty
\frac{d\tau}{\tau} f_k (\tau k^2)
\int \frac{d^4 q}{(2\pi)^4} tr_N \left\{ e^{-\tau (q^2 +
\frac{\partial^2 V_0}{\partial \phi_i \partial \phi_j} )}
- e^{-\tau (q^2 + g^2 \tilde{\vec{\phi^2}} )} \right\}\quad.
\ea

Because of the form of  $f_k (\tau k^2)$
derivatives of the effective potential with respect to $k$
are now infrared 
as well as ultraviolet regularized, so 
a further ultraviolet cutoff $\Lambda$ is no longer  necessary.
In the $\sigma$-model the second derivatives of the potential 
$V_0$ evaluated at $\tilde{\phi^2}$ are given by
\ba
\frac{\partial^2 V_0}{\partial \phi_i \partial \phi_j} 
& = & 
(-\mu_0^2 + \lambda_0 \tilde{\vec{\phi^2}} ) \delta_{ij} + 2
\lambda_0 \tilde{\phi_i} \tilde{\phi_j}\quad,\nonumber\\
& = &
\lambda_0 (\tilde{\vec{\phi^2}} - \phi_{k_0}^2) \delta_{ij} + 2
\lambda_0 \tilde{\phi_i} \tilde{\phi_j}
\ea
and the trace can explicitly be evaluated  for $N=4$ components.
We find for the trace
\ba\label{tracemaple}
tr_N e^{-\tau \frac{\partial^2 V_k}{\partial \phi_i \partial \phi_j} }
& = & 3 e^{-\tau \lambda_0 (\tilde{\vec{\phi^2}} - \phi_{k_0}^2)}
+ e^{-\tau \lambda_0 (3 \tilde{\vec{\phi^2}} - \phi_{k_0}^2)}\quad.
\ea
The first term on the r.h.s.\ of eq. (\ref{tracemaple}) 
corresponds to the eigenvalues associated
with the three pions while the other term describes the $\sigma$-meson.
Since the fermionic trace is diagonal in color-, flavor- and spin-spaces
we get
\ba
tr e^{-\tau g^2 \tilde{\vec{\phi^2}} } = 4 N_c N_f e^{- \tau g^2 
\tilde{\vec{\phi^2}} }\quad.
\ea
Using these results we 
are able to calculate the derivatives of $V_k(\phi)$ which we need 
for the renormalization group flow equations.

\section{Renormalization group flow equations}

We want to study the evolution of different quantities with respect
to the virtuality or momentum scale $k$. Since we are dealing with 
the possibility that the chiral symmetry is spontaneously broken 
we have to distinguish between two
phases. For large $k$ we start in the symmetric phase which is
defined by a vanishing vacuum expectation value (VEV) ($\phi_k =0$)
of the potential. Thus by means of the derivative with respect to
the fields
\ba
V ' & := & \frac{\partial V }{\partial \tilde{\phi^2}}
\ea
we can define the mass $m^2_k$ and the coupling constant $\lambda_k$
for the symmetric regime by
\ba
\frac{m^2_k}{2} & := & V ' ( \tilde{\phi^2}=\phi^2_k =0 )\quad,\nonumber\\
\frac{\lambda_k}{2}  & := & V '' ( \tilde{\phi^2}=\phi^2_k =0 )\quad.
\ea
Taking the derivative with respect to the scale $k$ we find
the following coupled sets of flow equations for the  
symmetric phase ($\phi^2_k = 0$):
\ba\label{eq22}
\frac{k}{2} \frac{\partial m^2_k}{\partial k} & = & k \frac{\partial V ' (0)}
{\partial k}\quad,\\
\label{eq23} 
\frac{k}{2} \frac{\partial \lambda_k}{\partial k} & = & k \frac{\partial V '' (0)}
{\partial k}\quad.
\ea

In the spontaneously broken region the VEV is finite ($\phi_k \neq 0$)
and the mass parameter $m^2_k$ tends to negative values. 
We prefer to parameterize the evolution of the potential in this
region in terms of $\lambda_k$ and the minimum of the potential 
$\phi_k$, which is defined by
\ba
V ' ( \phi_k ) &=& 0\quad.
\ea
This equation enables us to find the evolution of the minimum $\phi_k$.
By again taking the derivative with respect to the scale $k$ we find
the flow equation for $\phi^2_k$ in the
broken phase ($\phi^2_k \neq 0$).
Note that we first evaluate the derivatives on the right hand
sides for constant couplings with respect to $k$ and 
then improve this one-loop expression by substituting 
the running couplings $\lambda_k$, $m^2_k$ and the VEV 
$\phi^2_{k}$ instead of of $\lambda_0$, $m^2_0$ and the VEV 
$\phi^2_{k_0}$.
This procedure corresponds to  renormalization group improvement.
Thus we find the general flow equations for the broken phase
\ba
\frac{k}{2} \frac{\partial \phi^2_k}{\partial k} & = &
- \frac{k}{2 V '' (\phi^2_k )} \frac{\partial V ' (\phi^2_k )}{\partial k}
= - \frac{k}{\lambda_k} \frac{\partial V ' (\phi^2_k )}{\partial k}\quad,\\ 
\frac{k}{2} \frac{\partial \lambda_k}{\partial k} & = & 
k \frac{\partial V '' (\phi^2_k )}
{\partial k}\quad,\qquad \mbox{if}\quad V ''' = 0\quad.
\ea

\subsection{Evolution for  $T=0$}

The
heat-kernel representation of the effective potential yields 
the explicit evolution equations at $T=0$.  
In the symmetric phase at zero temperature we use the
equations (\ref{tracemaple},\ref{eq22},\ref{eq23}) and get:

\ba
\frac{k}{2} \frac{\partial m^2_k}{\partial k} & = &
-\frac{3 \lambda_k k^2}{(4 \pi)^2} 
\frac{1}{\left( 1 +  m^2_k/k^2 \right)^2}
+ \frac{4 N_c g^2 k^2}{(4 \pi)^2}\quad,\\
\label{lamsym}
\frac{k}{2} \frac{\partial \lambda_k}{\partial k} & = &
\frac{12 \lambda_k^2}{(4 \pi)^2} 
\frac{1}{\left( 1 + m^2_k/k^2 \right)^3}
- \frac{8 N_c g^4}{(4 \pi)^2}\qquad.
\ea

\begin{figure}[hbt]
\unitlength1cm
\begin{center}
\begin{picture}(15,9)(-1,-0.5)
\put(4,5){$\lambda_k$}
\put(5.5,-0.5){k [MeV]}
\epsfbox{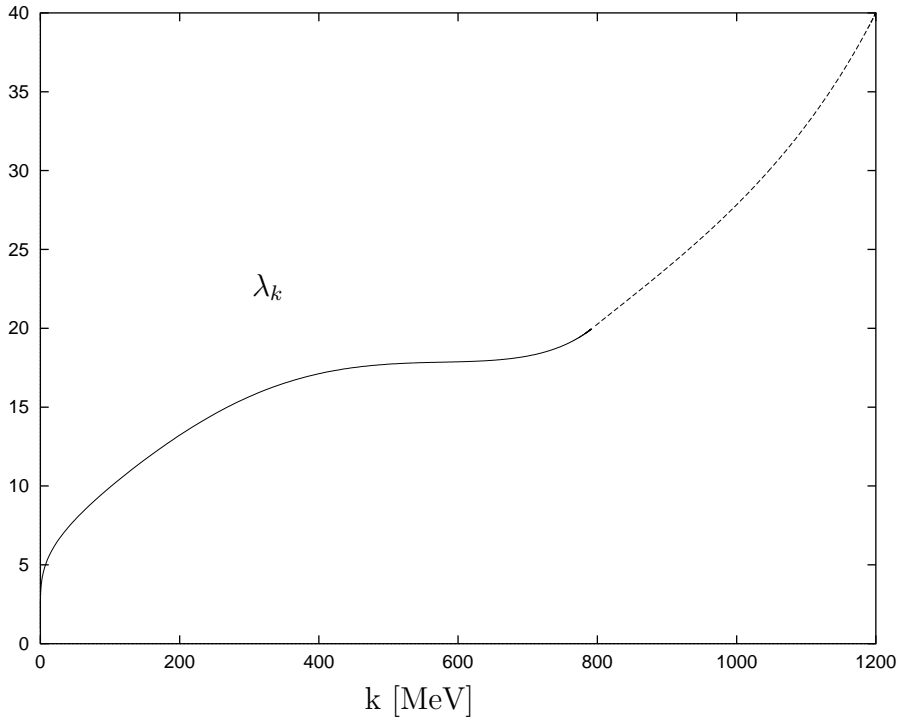}
\end{picture}
\parbox{12cm}
{\caption{\label{fig1} The evolution of $\lambda_k$ with 
respect to the scale $k$.}}
\end{center}
\end{figure}

Note the Yukawa coupling $g$ is fixed. 
In general we follow the evolution equations 
starting at $k_0 = 1.2$ GeV with the initial values
$\lambda_{k_0} = \lambda_0$ and $m^2_{k_0} = m^2_0$.
We proceed from large $k$ to small $k$ integrating out
more and more infrared modes. Near $k = k_0$ the
evolution of $\lambda_k$ is dominated by the bosonic
term proportional to $\lambda^2_k$ whereas the evolution of
$m^2_k$ responds to the fermion loop.
We reach chiral symmetry breaking at the scale  $k= \Lambda_{
\chi SB} \approx 0.8$
GeV, where we switch to the equations for the broken phase
characterized by $\phi_k^2 \neq 0$.  

\begin{figure}[hbt]
\unitlength1cm
\begin{center}
\begin{picture}(15,9)(-1,-0.5)
\put(5.5,-0.5){k [MeV]}
\put(3,2.5){$\phi_k$}
\put(9,6){$m_k$}
\put(-0.5,8){[MeV]}
\epsfbox{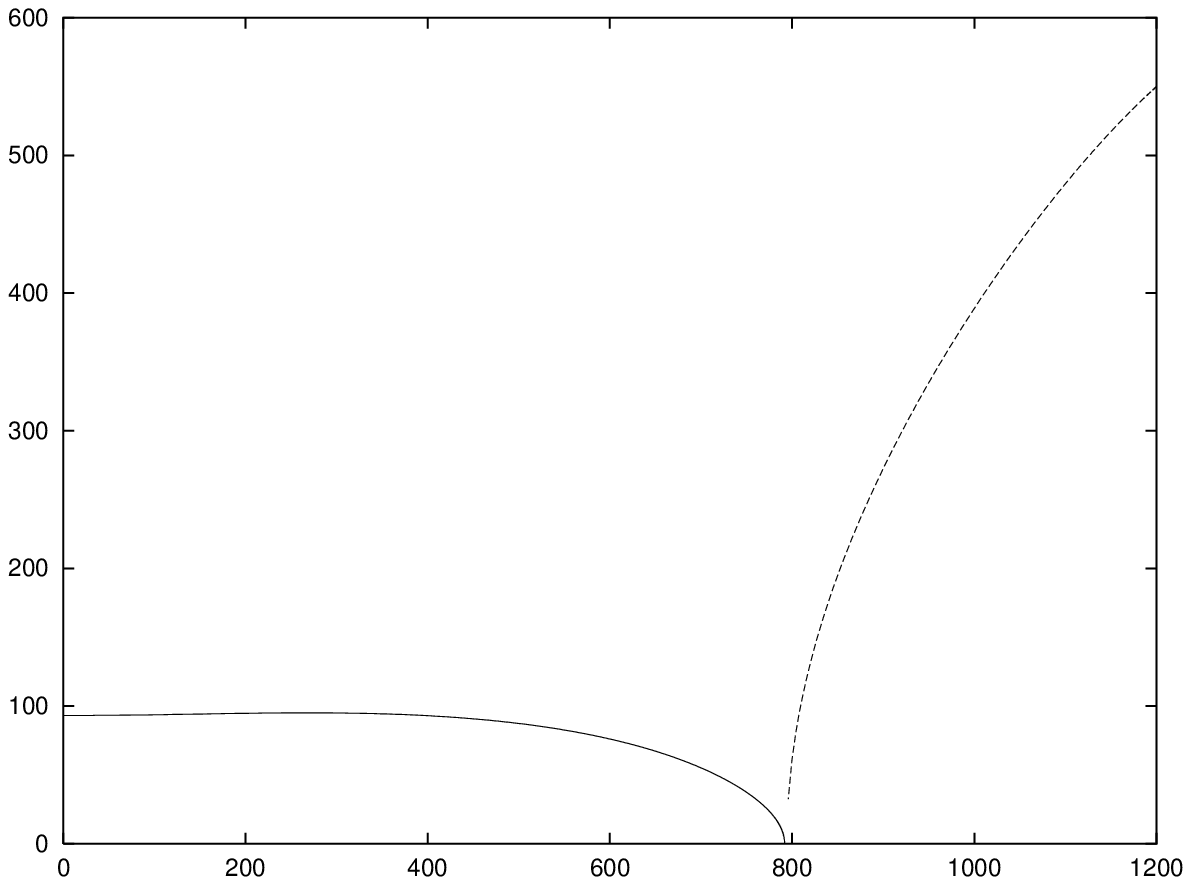}
\end{picture}
\parbox{12cm}
{\caption{\label{fig2} The evolution of the minimum 
of the effective potential.}}
\end{center}
\end{figure}

The squared mass of the $\sigma$-meson is then given
by $2 \lambda_k \phi_k^2$ and the quark mass by $g\phi_k$
and the evolution equations have the following form:

\ba
\frac{k}{2} \frac{\partial \phi^2_k}{\partial k} & = &
\frac{3 k^2}{2 (4 \pi)^2} \left[ 1 + 
\frac{1}{\left(1+2\lambda_k \phi^2_k / k^2\right)^2} \right]
- \frac{4 N_c}{(4\pi)^2}
\frac{k^2g^2}{\lambda_k}
\left[\frac{1}{\left(1+g^2 \phi^2_k / k^2\right)^2}\right]\ ,\\
\label{lambrok}
\frac{k}{2} \frac{\partial \lambda_k}{\partial k} & = &
\frac{3 \lambda_k^2}{(4 \pi)^2} \left[ 1 +
\frac{3}{\left(1+2\lambda_k \phi^2_k / k^2\right)^3} \right]
- \frac{8 N_c}{(4\pi)^2} g^4
\left[
\frac{1}{\left(1+g^2 \phi^2_k / k^2\right)^3}\right]\quad.
\ea
The first terms on the r.h.s.\ of the above equations
are related to the bosonic while the
last parts are connected to the fermionic contributions. 
The factor one in the first part comes from 
the three pions which are massless Goldstone
bosons while
the second part
of the bosonic term contains the contribution
of the $\sigma$-meson.
 
The $\sigma$-meson decouples from the evolution  when 
the term proportional to $2\lambda_k \phi^2_k/$ $k^2$
becomes large, i.e.\ when the $\sigma$-meson mass over the infrared
parameter $k$ is big.
(cf. \cite{wett}). The functions in squared brackets can be
called threshold functions, since they describe the decoupling 
of the pions and $\sigma$-mesons from the evolution.  
With the heat-kernel method and the choice of the function $f_k
(\tau)$ these threshold functions can be obtained
analytically. This is an advantage of our method compared
to the momentum cutoff used in refs. \cite{wett,dirk}. 
We remark that in the heat-kernel for the
effective potential the inverse fermion propagator enters
quadratically in the combination $D D^+$ therefore we can use the same
infrared cutoff function $f_k$ as in the bosonic integral 
without breaking chiral symmetry. If we want to evaluate
the running coupling more care has to be used to regulate
the fermion integration. 
One also recognizes the signs of the bosonic and fermionic
contributions to the $\beta$-function. The bosons lead to an infrared
stable (ultraviolet unstable) coupling, whereas the fermions counteract this
tendency. Going from high $k$ to low $k$ one sees that the 
mesonic self-interaction $\lambda_k$ will balance at intermediate values
of $k$, whereas in the far infrared the boson term wins 
(cf.\ figure~\ref{fig1}). At $k=k_{\chi SB} \approx 800$ MeV
eqs.(\ref{lamsym}) and (\ref{lambrok}) become identical, also visible in
figure~\ref{fig1} and the $\beta$-function of
$\lambda_k$ is continuous.
The vacuum expectation value $\phi_k$ stabilizes at small
values of $k$ and the evolution ends with $\lim_{k \to 0} \phi_k = f_\pi\quad$.
When the heavy particles, the sigma and quarks, have
decoupled, the change of the vacuum expectation value becomes proportional to
k which vanishes (see figure~\ref{fig2}).
The infrared $k_{\chi SB}$ scale found in this calculation is
somewhat smaller than the resolution $Q$ of the photon found in an
analysis of deep inelastic scattering \cite{pirn} for the transition 
from the quark as a parton to the massive constituent quark.
One has to wait for a more sophisticated calculation with running
Yukawa coupling $g_k$ and wave function renormalization $Z_k$
\cite{hdgr} which may improve the already astonishing agreement 
between phenomenology and the field theoretic model.

The above equations are the evolution equations for the running of the 
expectation value $\phi^2_k$
and quartic coupling $\lambda_k$ with the infrared scale $k$
at zero temperature.

\subsection{Evolution for finite $T$}

At finite temperature we integrate the momenta in equation (\ref{potential}),
by splitting the zero component from the three-dimensional spatial components
and convert the integration over $q_0$ into a summation over
Matsubara frequencies $\omega^2_n = 4\pi^2 n^2 T^2$ 
for the bosons and $\nu_n^2 = (2n+1)^2\pi^2T^2$ for the quarks.

In the symmetric phase the corresponding equations are written 
in terms of the positive mass parameter $m^2_k$ and $\lambda_k$:
\ba\label{symeq}
\frac{k}{2} \frac{\partial m_k^2}{\partial k} & = &
-\frac{3 \lambda_k^T}{(4 \pi)^2} k^2\left[
\frac{3\pi}{2} \frac{T}{k}
\sum_{n = -\infty}^\infty
\frac{1}{\left( 1 + \left(\omega^2_n+m^2_k\right)/{k^2} 
\right)^{5/2}}\right]\nonumber\\
&&\nonumber\\
&& + \frac{4 N_c g^2}{(4\pi)^2} k^2\left[\frac{3\pi}{2}\frac{T}{k}
\sum_{n = -\infty}^\infty
\frac{1}{\left( 1 + \nu^2_n/k^2 \right)^{5/2}}\right]\quad,\\
&&\nonumber\\ 
\frac{k}{2} \frac{\partial \lambda_k^T}{\partial k} & = &
\frac{12 (\lambda_k^T)^2}{(4 \pi)^2}
\left[\frac{15\pi}{8}\frac{T}{k}
\sum_{n = -\infty}^\infty
\frac{1}{\left( 1 + \left(\omega^2_n+m^2_k\right)/{k^2} 
\right)^{7/2}}\right]\nonumber\\
&&\nonumber\\
&& -\frac{8 N_c g^4}{(4\pi)^2}
\left[\frac{15\pi}{8}\frac{T}{k}
\sum_{n = -\infty}^\infty
\frac{1}{\left( 1 + \nu^2_n/k^2 \right)^{7/2}}\right]\quad.
\ea

\begin{figure}[hbt]
\unitlength1cm
\begin{center}
\begin{picture}(15,9)(-1,-0.5)
\put(5.5,-0.5){T/k}
\put(6,5){$y=0$}
\put(8,3.5){$y=2$}
\put(9,2.5){$y=4$}
\put(2,8){$\frac{15 \pi}{8}\frac{T}{k}
\sum\limits_{n = -\infty}^\infty
\frac{1}{\left( 1 + \omega^2_n/k^2 + y^2\right)^{7/2}}$}
\epsfbox{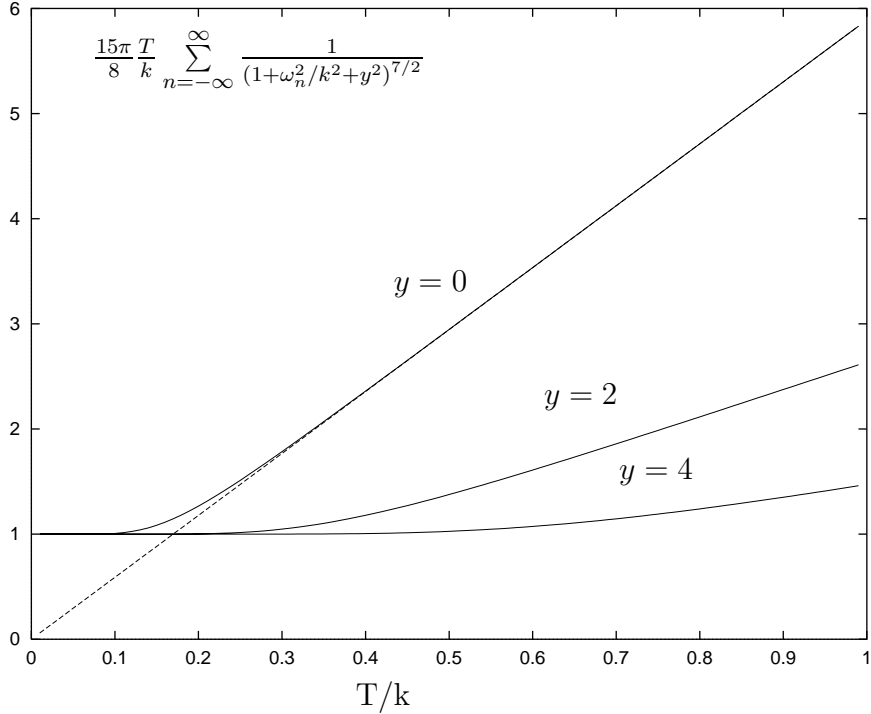}
\end{picture}
\parbox{12cm}
{\caption{\label{fig3} The bosonic threshold functions for 
different mass parameters $y=0,2,4$ as function of $T/k$. 
The dashed line is the function $15 \pi /8\cdot T/k$ which 
demonstrate the linear behavior of the threshold function for
large $T/k$.}}
\end{center}
\end{figure}

In the broken phase one finds:
\ba
\label{brokeneq}
\frac{k}{2} \frac{\partial (\phi^T_k)^2}{\partial k} & = &
\frac{3 k^2}{2(4 \pi)^2}\left[\frac{3\pi}{2}\frac{T}{k}
\sum_{n = -\infty}^\infty\left\{
\frac{1}{\left( 1 + \omega^2_n/k^2 \right)^{5/2}} +
\frac{1}
{\left( 1 + 
\left(\omega^2_n + 2\lambda^T_k (\phi^T_k)^2\right)/{k^2}\right)^{5/2}}
\right\}
\right]\nonumber\\
&&\nonumber\\
&& -\frac{4 N_c}{(4\pi)^2} \frac{g^2 k^2}{\lambda^T_k}
\left[\frac{3\pi}{2}\frac{T}{k}
\sum_{n = -\infty}^\infty
\frac{1}{\left( 1 + 
\left(\nu^2_n+ g^2 (\phi^T_k)^2\right)/{k^2} \right)^{5/2}}
\right]\quad,\\
&&\nonumber\\
\label{brokenlambda}
\frac{k}{2} \frac{\partial \lambda_k^T}{\partial k} & = &
\frac{3 (\lambda_k^T)^2}{(4\pi)^2}\left[
\frac{15 \pi}{8}\frac{T}{k}
\sum_{n = -\infty}^\infty
\left\{
\frac{1}{\left( 1 + \omega^2_n/k^2 \right)^{7/2}} +
\frac{3}
{\left( 1 + 
\left(\omega^2_n + 2\lambda^T_k (\phi^T_k)^2\right)/{k^2}\right)^{7/2}} 
\right\}
\right]\nonumber\\
&&\nonumber\\
&& -\frac{8 N_c}{(4\pi)^2} g^4 \left[
\frac{15 \pi}{8}\frac{T}{k}
\sum_{n = -\infty}^\infty
\frac{1}{\left( 1 + 
\left(\nu^2_n + g^2 (\phi^T_k)^2\right)/{k^2} \right)^{7/2}}
\right]\quad.
\ea

\begin{figure}[hbt]
\unitlength1cm
\begin{center}
\begin{picture}(15,9)(-1,-0.5)
\put(5.5,-0.5){T/k}
\put(5,3){$y=0$}
\put(7,5.5){$y=2$}
\put(9,7){$y=4$}
\put(4,8){$\frac{15 \pi}{8}\frac{T}{k}
\sum\limits_{n = -\infty}^\infty
\frac{1}{\left( 1 + \nu^2_n/k^2 + y^2 \right)^{7/2}}$}
\epsfbox{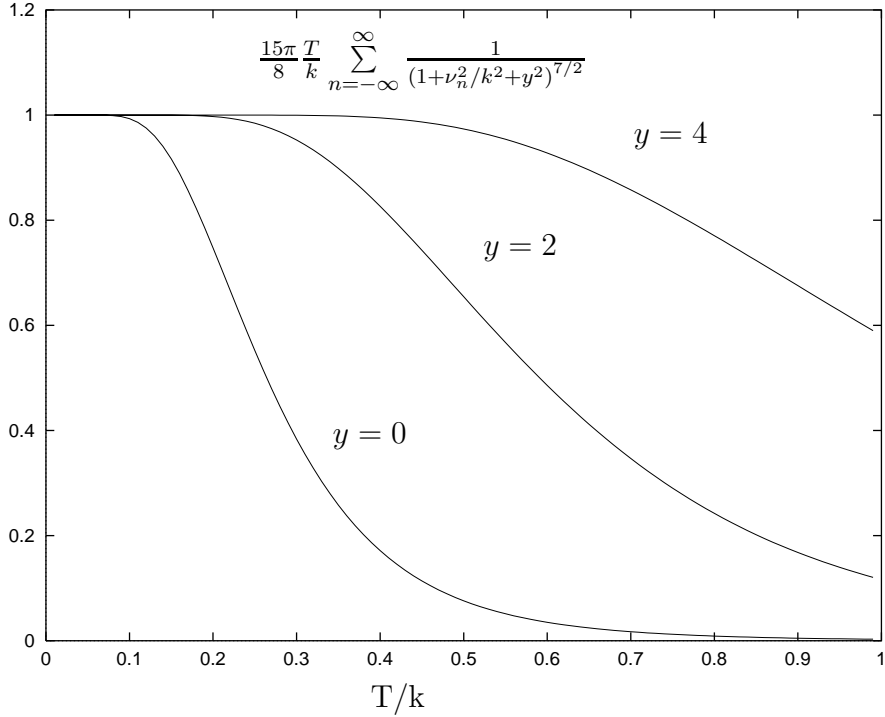}
\end{picture}
\parbox{12cm}
{\caption{\label{fig4} The fermionic threshold functions for different
mass parameters $y=0,2,4$ as function of $T/k$.}}
\end{center}
\end{figure}

One sees that the form of the equations is unchanged concerning the
dependence on the coupling constants. Only the threshold functions
given in square brackets are different at finite temperature.
Due to  the three dimensional
momentum integrations fractional powers arise in the threshold functions.

In the limit of low temperatures $T/k \to 0$ we 
regain our old expressions at $T=0$, i.e.\ we find the relations: 

\ba
\frac{3\pi}{2} \frac{T}{k}
\sum_{n = -\infty}^\infty
\frac{1}{\left( 1 + \left\{\omega^2_n \atop \nu^2_n \right\} 
/{k^2} +y^2\right)^{5/2}} 
& \to & \frac{1}{\left( 1 + y^2 \right)^2}
\ea
and also for the other threshold functions 
\ba\label{s1}
\frac{15 \pi}{8}\frac{T}{k}
\sum_{n = -\infty}^\infty
\frac{1}{\left( 1 +  \left\{\omega^2_n \atop \nu^2_n \right\}
/k^2 + y^2\right)^{7/2}} 
& \to & \frac{1}{\left( 1 + y^2 \right)^3}\quad.
\ea

These equations guarantee the right matching 
of the finite temperature equations
to the zero temperature equations, i.e.\ in the limit
$T \to 0$
the set of equations for finite and zero temperature become identical.
For large ratios $T/k$ the 
bosonic threshold functions, given in the square brackets 
of eqs. (\ref{symeq})-(\ref{brokenlambda}),
increase linearly in $T/k$. (cf.\ figure~\ref{fig3}).

This increase is due to the $n=0$ Matsubara mode in the frequency sum.
In the large temperature limit the ($3+1$)-dimensional system reduces
to a $3$-dimensional system. 
We plot in figure~\ref{fig3} the threshold functions for the bosons
and in figure~\ref{fig4} the threshold functions for the fermions.
The dimensional reduction can also be seen in
in the critical behavior, as will be shown in the next section.
The fermionic threshold functions decrease with $T/k$ and
run to zero. There also exists a stable plateau in the vicinity of the
origin for the Fermi-case as in the ref.~\cite{dirk}. 

\section{Numerical results and discussion}

The procedure to obtain finite temperature results is now the following.
For each fixed temperature $T$ we solve the evolution equations
as functions of $k$
with the same starting parameters as the $T=0$ theory.
The underlying 
idea is that at the large ultraviolet scale the finite temperature
does not modify the effective theory, since 
the finite temperature only modifies the
boundary in the imaginary time direction. 

\begin{table}[h]
\begin{center}
\begin{tabular}{cc||ccl} 
$\lambda_{k_0}$ & 
$m_{k_0}\ [\mbox{MeV}]$ & 
$k_{\chi SB}\ [\mbox{MeV}] $ & 
$f_{\pi}^{T=0}\ [\mbox{MeV}]$ &
$\quad T_c\ [\mbox{MeV}]$\\[0.5ex] \hline\hline
30 & 600 & 787 & 94 & $\simeq 132 \pm 1$ \\ 
40 & 550 & 790 & 93 & $\simeq 131.5 \pm 0.1$ \\ 
60 & 450 & 803 & 93 & $\simeq 133 \pm 1$ \\ 
90 & 300 & 815 & 93 & $\simeq 133 \pm 1$ \\ 
120 & 70 & 823 & 93 & $\simeq 134 \pm 1$ \\ \hline 
\end{tabular}
\parbox{12cm}
{\caption{\label{tab1} Different initial values compared with 
the corresponding critical temperatures.}}
\end{center}
\end{table}

\begin{figure}[hbt]
\unitlength1cm
\begin{center}
\begin{picture}(15,9)(-1,-0.5)
\put(6,2.5){$\phi_k (T)$}
\put(9,6){$m_k (T)$}
\put(-0.5,8){[MeV]}
\put(2,2){$T = 11$ MeV}
\put(4.2,0.8){$T=131$ MeV}
\put(5.5,-0.5){k [MeV]}
\epsfbox{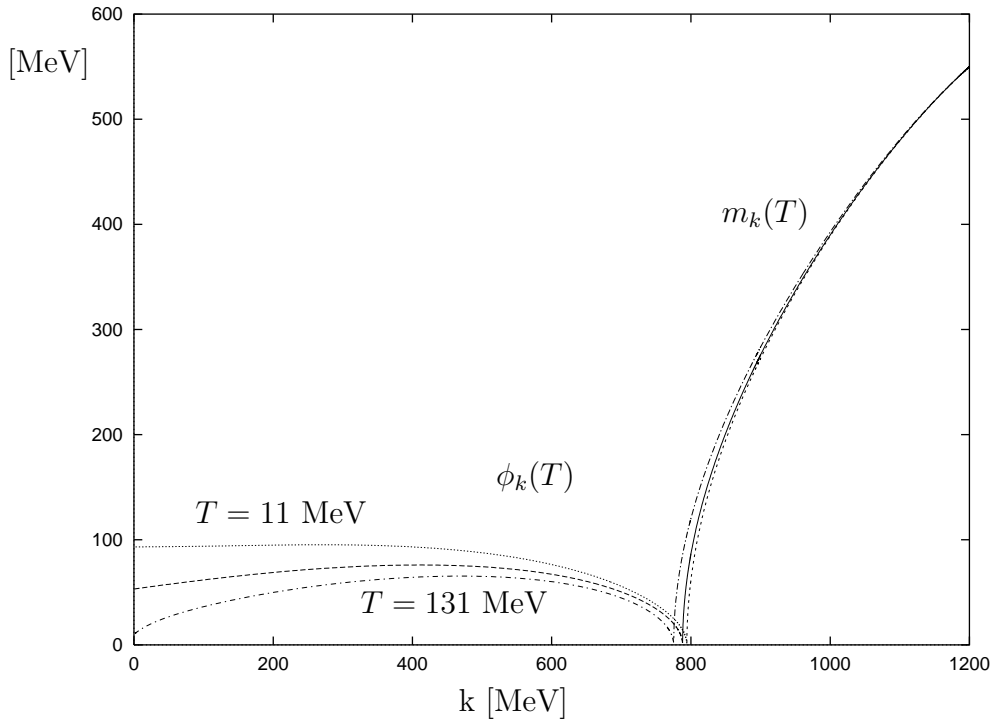}
\end{picture}
\parbox{12cm}
{\caption{\label{fig5} VEV $\phi_k$ and mass $m_k$ as 
function of $k$ for different temperatures. (Upper line $T=11$ MeV, middle
line $T=111$ MeV and bottom line $T=131$ MeV).}}
\end{center}
\end{figure}

At the initial cutoff scale
the momenta at the ultraviolet cutoff are anyhow so large that the  theory 
is not affected
by temperatures $2 \pi T < 1.2$ GeV. 
Of course, for higher temperatures
also the  input would have to be modified.  
With increasing temperature the mass parameter $m_k$
decreases slightly 
more slowly towards the condensation point (cf.\ figure~\ref{fig5}),
but the main effect of the finite temperature occurs below 
$k\simeq 800$ MeV.

The threshold functions contain an additional damping due to the
Matsubara frequency. Below the condensation point the
boson condensate $\phi_k^2$ reaches less high  values at finite 
temperatures than before 
at $T=0$. For all temperatures it decreases again with $k \to 0$, which 
is due to the pion loop.
After optimizing the Runge-Kutta code a critical 
temperature 
\ba
T_c \approx 131,5\  \mbox{MeV}
\ea 
was found.

This temperature would be compatible with that of the Wetterich group
if $f_\pi= 93$ MeV had been used in \cite{dirk} even in the chiral
limit \cite{priv}. They obtained a critical temperature 
$T_c \approx 100$ MeV by using non-analytic threshold functions and
wave function and coupling constant renormalization.
Both approaches share the linear sigma model with free quarks even at low
temperature. In table~\ref{tab1} other possible initial 
values for $\lambda_{k_0}$ and $m_{k_0}$  at the ultraviolet scale 
are shown which would
give the same $f_{\pi}$ value for vanishing temperature.
The corresponding critical temperature and 
$k_{\chi SB}$ scale are almost constant and are correlated like 
\be
2 \pi T_c \approx k_{\chi SB}\quad.
\ee

\begin{figure}[hbt]
\unitlength1cm
\begin{center}
\begin{picture}(15,9)(-1,-0.5)
\put(-0.8,8){$\frac{\D \phi_{k=0} (T)}{\D \phi_{k=0} (0)}$}
\put(-0.5,7.2){[MeV]}
\put(5.5,-0.5){Temperature [MeV]}
\put(8.5,8){chiral pert.\ theory}
\put(5.9,3.9){$\left(\frac{\D T_c -T}{\D T_c}\right)^\beta$}
\epsfbox{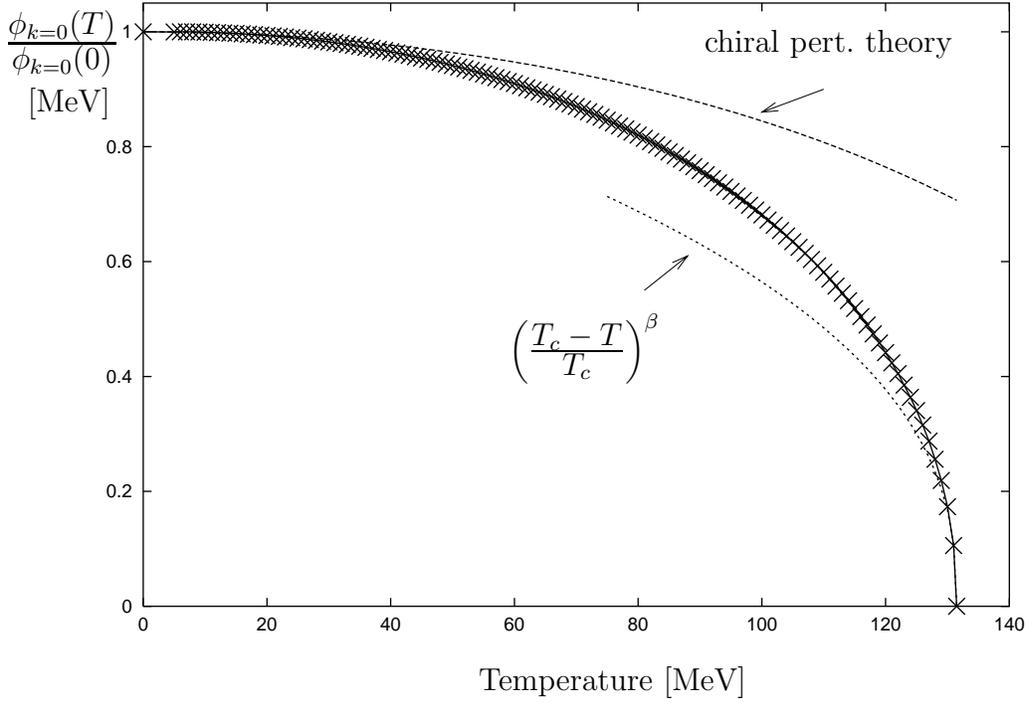}
\end{picture}
\parbox{12cm}
{\caption{\label{fig6} The order parameter as function of $T$.}}
\end{center}
\end{figure}

\begin{figure}[hbt]
\unitlength1cm
\begin{center}
\begin{picture}(15,9)(-1,-0.5)
\put(-1.,8){$log \ \frac{\D \phi_k}{\D \mbox{[MeV]}} $}
\put(5.5,-0.5){$log \ \frac{\D T_c - T}{\D \mbox{[MeV]}} $}
\put(2,8){Critical Exponent $\beta$}
\put(7.8,2.6){mean field}
\put(6.9,6.4){$\beta = 0.40$}
\epsfbox{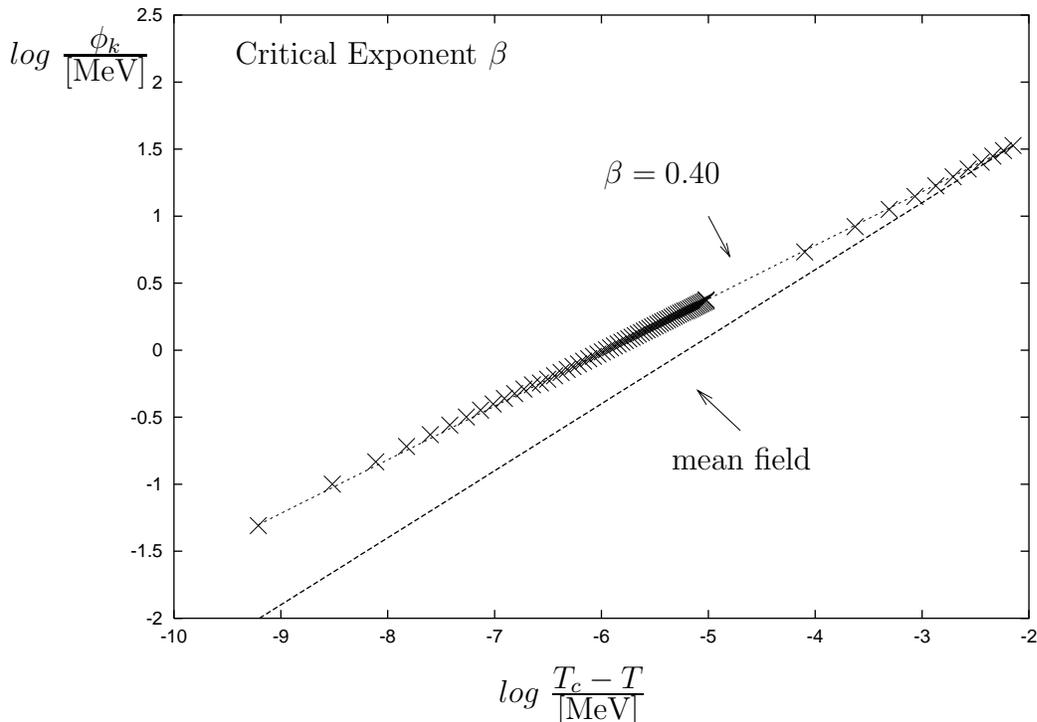}
\end{picture}
\parbox{12cm}
{\caption{\label{fig7} The critical exponent $\beta$.}}
\end{center}
\end{figure}

In figure~\ref{fig6} we show the normalized order parameter 
$\phi_{k=0} (T)/\phi_{k=0} (0)$ 
as a function of temperature. It follows the result of
chiral perturbation theory until $T \approx 35$ MeV, then it
deviates because of the stronger effects of the quark
loops. 
In chiral perturbation theory the temperature dependence of the 
light quark condensate for massless
quarks, which is plotted in figure~\ref{fig6}, is given by the
expression
\ba
\frac{\langle \bar{q} q\rangle_T}{\langle \bar{q} q\rangle_0}
& = & 1-\frac{T^2}{8 f_\pi^2} -\frac{T^4}{384 f_\pi^4}-
\frac{T^6}{288 f_\pi^6}\ln \frac{\Lambda_q}{T} + {\cal O}(T^8)
\ea
with $\Lambda_q = 470 \pm 110$ MeV \cite{chpt}.
The purely mesonic description of the phase transition
with a finite number of mesons
becomes inadequate since more and more mesons become important.
In the vicinity of $T_c$ we also extrapolated in figure~\ref{fig6} 
the scaling behavior of the order parameter with our 
critical exponent $\beta \approx 0.40$.
Still in a model without confinement of quarks the fermion
loop may be overestimated cf.~\cite{wachs}.  
The renormalization group allows to include the long range
correlations properly near the critical point. 
In fact  it is possible to see scale invariance by analyzing 
the power
law behavior of the order parameter near $T_c$.
We plot $log (\phi_k)$ versus 
$log (T_c - T)$ in figure~\ref{fig7}. 
The data points lie on a linear curve given by
\ba
0.40 log(T_c - T) + (1-0.40) log(T_c) - 0.30\quad.
\ea
This yields a critical exponent $\beta \approx 0.40$ which is 
in good agreement 
with lattice results for the $O(4)$-theory
in three dimensions \cite{mc}. 
So in spite of the finite temperature the limit $T/k \to \infty$ 
enforces the dimensional reduction. One knows, that 
the linear sigma model lies  in the same universality class
as the $O(4)$-theory, since the fermions are not contributing 
to the critical fluctuations near $T_c$ \cite{pis}.
To compare the result with the mean field value $\beta = 0.5$ we also
plotted this slope in the figure~\ref{fig7}.
Further investigations must study how far away from the critical point
the renormalization group improvement is important for the behavior 
of the order parameter. 
An especially interesting point is to find the window for the
critical dynamics. In the superconducting phase transition only the
mean field behavior is experimentally relevant. From the figure~\ref{fig6} 
it seems that in chiral QCD the critical behavior influences a wider
range of temperatures. In real QCD, however, the finite quark
masses spoil the second order phase transition. Also gluon effects will
change the behavior of the pressure in comparison with the effective
linear $\sigma$-model. 
Numerical simulations of lattice QCD still show an inconclusive 
critical behavior of the order parameter. 
In lattice QCD the main problem is the 
treatment of light particles leading to large correlation lengths which
are larger than the lattice size.

The presented method of heat-kernel regularization opens the way for more
refined calculations also including coupling constant and wave function
renormalization. Here the more delicate problem of introducing a cutoff
which does not violate chiral symmetry has to be dealt with.  Besides 
comparing the $T=0$ theory to the behavior of deep inelastic scattering
we plan to investigate the spectrum of the Dirac operator
\cite{hdgr}. 
Since our method is a multi-scale analysis of the effective theory
we think that the Dirac eigenmodes up to $k=1.2$ GeV can be checked
and compared to lattice calculations. 
Random matrix theory has been extremely successful
to predict the smallest eigenvalues and correlations of nearest neighbor
levels. Our method should give the broad behavior of the Dirac spectrum
at $T=0$ and for finite temperatures. In our method the real dynamics 
of QCD at low resolution seems to be captured quite well. So we are optimistic
about further prospects.
These include the discussion of three flavor dynamics and finite 
quark mass effects \cite{dagr}. Due to the simplicity of our cutoff function
we expect also a considerable simplification in the set up and solution
of three flavor dynamics.
For nuclear physicist remains the ever challenging field of finite
baryon density where analytical methods like our renormalization
group equations are very promising to model the transition of
the nuclear many
body system to the quark many body system.

\section*{Acknowledgment}

B.-J.~S. would like to thank the organizers of the workshop, in
particular D.~Blaschke, for the invitation and warm hospitality.
We (B.-J.~S. and H.J.P.) also thank C.~Wetterich and D.-U. Jungnickel for many 
fruitful and helpful discussions.


\begin{thebibliography}{MMMM}

\bibitem{wett} C.~Wetterich, \PL{B301} (1993) 90; 
  C.~Wetterich and N.~Tetradis, \IJMP{A9} (1994) 4029.
\bibitem{block} K.~G.~Wilson, \PR{B4} (1971) 3174; K.~G.~Wilson and
  I.~G.~Kogut, \PRP{12} (1974) 75; F. Wegner and A. Houghton, \PR{A8}
  (1973) 401;
  F.~Wegner in {\em Phase Transitions and Critical Phenomena}, vol.~6,
  eds.~C. Domb and M.S. Greene, Academic Press (1976); J.F.  Nicoll
  and T.S. Chang, \PL{62A} (1977) 287; 
  S.~Weinberg in {\em Critical Phenomena for Field Theorists}, 
  Erice Subnucl.~Phys. (1976) 1;
\bibitem{polc} J.~Polchinski, \NP{B231} (1984) 269;
\bibitem{dirk} D.-U.~Jungnickel and C.~Wetterich, \PR{D53} (1996)
  5142, {\tt hep-ph/9505267}; J.~Berges, D.-U.~Jungnickel and C. Wetterich, 
  {\em Two Flavor Chiral Phase Transition from Nonperturbative Flow 
    Equations}\/(1997), {\tt hep-ph/9705474}.
\bibitem{pirn} H. G. Dosch, T. Gousset and H. J. Pirner,
  {\em Nonperturbative $\gamma p^*$ Interaction in the Diffractive regime},
  to be published in Phys. Rev. D (1997), {\tt hep-ph/9707264}.
\bibitem{flor} R.~Floreanini and R.~Percacci, \PL{B356} (1995) 205.
\bibitem{bjhj} B.-J.~Schaefer and H.~J.~Pirner, 
  {\em Application of the heat-kernel method at finite temperature}, 
  to be published in Nucl. Phys. A (1997), {\tt hep-ph/9706258}.
\bibitem{mc} K.~Kanaya and S.~Kaya, \PR{D51} (1995) 2404.
\bibitem{morri} T.R.~Morris and M.D.~Turner, {\tt hep-th/9704202}. 
\bibitem{priv} J.~Berges and D.-U.~Jungnickel, private communication.
\bibitem{eps} G.~Baker, D.~Meiron and B.~Nickel, \PR{B17} (1978)
  1365.
\bibitem{eguc} T.~Eguchi, \PR{D14} (1976) 2755; 
  S.~P.~Klevansky, \RMP{64} (1992) 649.
\bibitem{eber} D.~Ebert, Th.~Feldmann and H.~Reinhardt, \PL{B388}
  (1996) 154; D.~Ebert and H.~Reinhardt, \NP{A271} (1986) 188.

\bibitem{pis} R.~D. Pisarski and F.~Wilczek, \PR{D29} (1984) 338.
\bibitem{chpt} J.~Gasser and H.~Leutwyler, \PL{B184} (1987) 83; 
  P.~Gerber and H.~Leutwyler, \NP{B321} (1989) 387; D.~Toublan,
  {\em Pion Dynamics at Finite Temperature}, {\tt hep-ph/9706273}. 
\bibitem{wachs} H.~J. Pirner and M. Wachs, \NP{A617} (1997) 395.
\bibitem{hdgr} G.~Papp, H.~J. Pirner and B.-J.~Schaefer, work in progress.
\bibitem{dagr} Z.~Aouissat, B.-J.~Schaefer and J.~Wambach,
  work in progress. 
\end{thebibliography}
\end{document}